\documentclass{article}
\usepackage{spconf,amsmath,graphicx,cite}

\usepackage{enumerate}

\usepackage{amsfonts,amssymb}
\usepackage[mathscr]{eucal}

\usepackage{color}
\definecolor{red}{RGB}{0,0,0}

\DeclareMathOperator{\diag}{diag}

\allowdisplaybreaks

\newcommand{\tenseur}[1]{\boldsymbol{\mathscr{#1}}}
\newcommand{\out}{\otimes}
\newcommand{\kron}{\boxtimes}
\newcommand{\khatri}{\odot}
\newcommand{\vect}{\text{vec}}
\newcommand{\unvect}{\text{unvec}}
\newtheorem{proposition}{Proposition}[section]

\definecolor{brickred}{rgb}{0.6,0,0}
\definecolor{byzantium}{rgb}{0.44,0.16,0.6}
\definecolor{darkgreen}{rgb}{0,0.6,0}

\title{Statistical efficiency of structured CPD estimation applied to Wiener-Hammerstein modeling}
\name{Jos\'{e} Henrique de M. Goulart$^{*,\mathsection}$, Maxime Boizard$^{**,\dagger}$, R\'{e}my Boyer$^{**}$, G\'{e}rard Favier$^{*}$
and Pierre Comon$^\ddagger$
\thanks{The authors wish to thank DIGITEO for its financial support on the project DETMOTS-2A.}
\thanks{$^\mathsection$ Sponsored by CNPq--Brazil (individual grant 245358/2012-9).}
\thanks{$^\ddagger$ Supported by ERC grant AdG-2013-320594 DECODA.}
}

\address{ %
\normalsize	$^*$ I3S, Universit\'{e} Nice Sophia Antipolis, CNRS, France\\
\normalsize	$^{**}$ L2S, Universit\'{e} Paris-Sud, CNRS, CentraleSupelec, France\\
\normalsize	$^{\dagger}$ SATIE, ENS Cachan, France\\
\normalsize	$^\ddagger$ GIPSA-Lab, Universit\'{e} de Grenoble, CNRS, France\\
}

\begin{document}

\maketitle
\begin{abstract}
 The computation of a structured canonical polyadic decomposition (CPD) is useful to address several important modeling problems in 
real-world applications. In this paper, we consider the identification of a nonlinear system by means of a Wiener-Hammerstein model, 
assuming a high-order Volterra kernel of that system has been previously estimated. Such a kernel, viewed as a tensor, admits a CPD 
with banded circulant factors which comprise the model parameters. To estimate them, we formulate specialized estimators based on 
recently proposed algorithms for the computation of structured CPDs. Then, considering the presence of additive white Gaussian noise, 
we derive a closed-form expression for the Cramer-Rao bound (CRB) associated with this estimation problem. Finally, we assess the 
statistical performance of the proposed estimators via Monte Carlo simulations, by comparing their mean-square error with the CRB.
\end{abstract}
\begin{keywords}\small
Tensor Decomposition, Structured CPD, Cram\'{e}r-Rao bound, Wiener-Hammerstein model
\end{keywords}
\vspace{-1ex}

\section{Introduction}
\label{sec:intro}

\vspace{-1.5ex}
The canonical polyadic decomposition (CPD), which can be seen as one possible extension of the SVD to higher-order tensors 
\cite{Comon2014},
is by now a well-established mathematical tool utilized in many scientific disciplines \cite{Kolda2009}. As it requires only mild
assumptions for being essentially unique, the CPD provides means for blindly and jointly identifying the components of multilinear
models, which arise in many real-world applications; see \cite{Comon2014, Kolda2009, Lim2014} for some examples.

In particular, the computation of CPDs having structured factors--such as Vandermonde, Toeplitz or Hankel matrices--has been shown
useful in problems including  channel estimation \cite{Fernandes2008}, nonlinear system identification \cite{Favier2009} and 
multidimensional
harmonic retrieval \cite{Sorensen2013b}.
As a consequence, several special-purpose algorithms have been  developed \cite{Kibangou2009, Sorensen2013b, Sorensen2013, 
Goulart2014}.
\\
\indent In practice, the data tensor to be decomposed is always corrupted by noise. Therefore, the assessment of the statistical 
performance of
CPD computation algorithms via comparison with the Cram\'{e}r-Rao bound (CRB) \cite{Kay1993} is of practical interest, since it can 
guide
the choice for an appropriate algorithm in application domains. Furthermore, it can provide valuable information for the study and
development of such algorithms. For the unstructured CPD, \cite{Liu2001} has derived the associated CRB and presented an evaluation of 
the
popular alternating least-squares (ALS) algorithm for tensors of orders three and four. Regarding the structured case, the CRB for the
estimation of a complex CPD with a particular Vandermonde factor has been derived in \cite{Sahnoun2014}, motivated by the problem of
estimating the directions of arrival of multiple source signals. Also, \cite{Boizard2015} has provided a closed-form expression
for the CRB associated with the estimation of a CPD having Hankel and/or Toeplitz factors.
\\
\indent This paper addresses the statistical evaluation of algorithms specialized in computing a CPD having banded circulant factors 
when
applied to estimate the parameters of a Wiener-Hammerstein (WH) model, which is a well-known block-oriented model used for representing
nonlinear dynamical systems \cite{Haber1999}. Because many systems of practical relevance can be (approximately) described by the WH 
model,
the problem of identifying its parameters from a set of experimental data (\textit{i.e.}, measured input and output samples) is 
well-studied; see, \textit{e.g.}, \cite{Haber1999} and references therein. One possible approach, as described in \cite{Favier2009}, 
consists in estimating the WH model parameters by computing the structured CPD of a kernel of its equivalent Volterra model. Here, we 
derive a closed-form expression for the CRB associated with this estimation problem, assuming the availability of a previously 
identified Volterra kernel corrupted by white Gaussian noise. Then, we formulate specialized estimation algorithms based on the CP 
Toeplitz (CPTOEP) and circulant-constrained ALS (CALS) methods proposed in \cite{Sorensen2013, Goulart2014} and evaluate their 
performance by comparing their mean-square error with the CRB through Monte Carlo simulations.

\pagebreak[3]

\textbf{Notation.}\label{sec:notation}
Scalars are denoted by lowercase letters, \textit{e.g.} $\theta_i$ or $a_{ij}$, vectors  by lowercase boldface, \textit{e.g.}
$\boldsymbol{\theta}$ or $\mathbf{a}_j$, matrices by  boldface capitals, \textit{e.g.} $\mathbf{B}$ or $\mathbf{A}^{(p)}$, and higher order
arrays by calligraphic letters, \textit{e.g.} $\tenseur{X}$.  We use the superscripts $^{T}$ for transposition, $^\dag$ for pseudo-inverse,
$\kron$ and $\khatri$ denote the Kronecker and Khatri-Rao products, respectively, and $\out$ stands for the (tensor) outer product. The
shorthand $\mathbf{a}^{\kron p}$ denotes $\mathbf{a} \kron \dots \kron \mathbf{a}$, where $\mathbf{a}$ appears $p$ times;
$\mathbf{a}^{\out p}$ and $\mathbf{A}^{\khatri p}$ are defined analogously. For our purposes, a tensor $\tenseur{X}$ of order $P$ will be
assimilated to its array of coordinates, which is indexed by $P$ indices. Its entries will be denoted by $x_{i_1, \ldots , i_P}$.

\section{Wiener-Hammerstein model identification via CPD}
\label{sec:background}

\subsection{Tensors and the CP decomposition}
\label{sec:tensors}

The polyadic decomposition of a $p$th-order tensor is defined by\vspace{-1.2ex}
\begin{equation}\vspace{-1ex}
\tenseur{X}= \sum_{r=1}^R \mathbf{a}_r^{(1)} \out \mathbf{a}_r^{(2)} \out  \ldots \out  \mathbf{a}_r^{(p)},
\label{eq:cp}
\end{equation}
where $\mathbf{a}_r^{({q})}$ is the $r^{th}$ column of $\mathbf{A}^{({q})} \in \mathbb{R}^{I_q \times R}$. The minimal value of $R$ 
such that $\tenseur{X}$ can be written as in (\ref{eq:cp}) is called the rank of $\tenseur{X}$, in which case we refer to the above 
decomposition as the CPD of $\tenseur{X}$. Another way of expressing \eqref{eq:cp} is
{\color{red}\begin{equation*}
 \tenseur{X} = \tenseur{I} \times_1 \mathbf{A}^{(1)} \times_2 \dots \times_p \mathbf{A}^{(p)},
 \label{CPD-moden}
\end{equation*}
where} $\tenseur{I} \in \mathbb{R}^{R \times \dots \times R}$ is {\color{red}a $p$th-order diagonal tensor such that 
$[\tenseur{I}]_{r,\ldots,r}=1$} and $\times_q$ denotes the {\color{red}mode-$q$} product (see, \textit{e.g.}, 
\cite[Sec.~2.5]{Kolda2009}).
  
\subsection{The WH model and its equivalent Volterra model}
\label{sec:wiener-hamm}

\begin{figure}[t!]
  \centering
      \begin{picture}(385,15)(-8,-15)

	  \put(0,-15){\vector(1,0){30}}
	  \put(5,-11){\footnotesize $u(n)$}
 	  \put(30,-22){\framebox(35, 14){\footnotesize $W(z)$}}	
	  \put(65,-15){\vector(1,0){25}}
	  \put(90,-22){\framebox(40, 14){\footnotesize $g(\cdot)$}}
 	  \put(130,-15){\vector(1,0){25}}
 	  \put(155,-22){\framebox(35, 14){\footnotesize $H(z)$}}
	  \put(196,-11){\footnotesize $y(n)$}
	  \put(190,-15){\vector(1,0){35}}
	
      \end{picture}
      \caption{Block-diagram  of the Wiener-Hammerstein model.}
      \label{fig:wiener-hamm}
\end{figure}
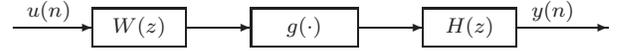

The structure of a discrete-time WH model is as depicted in Fig.~\ref{fig:wiener-hamm}. Basically, it consists of a cascade connection
comprising a memoryless nonlinearity $g(\cdot)$ ``sandwiched'' by two linear systems, $W(z)$ and $H(z)$. Because of its structured form
constituted by fundamental blocks, the WH model is said to belong to the class {\color{red}of} block-oriented models \cite{Haber1999}.

In this paper, we consider the time-invariant WH model constituted by a polynomial {\color{red} nonlinearity}  $g(x) = \sum_{p=1}^P g_p
x^p$ and {\color{red}by} finite impulse response filters $W(z) = \sum_{l=0}^{L_w-1} w_l z^{-l}$, {\color{red} with $w_0 \neq 
0$, and $H(z) = \sum_{r=0}^{R-1} h_r z^{-r}$. }
Hence, the resulting expression relating the input $u(n)$ to the output $y(n)$ is
\begin{equation}
  y(n) = \sum_{p=1}^P g_p \sum_{r=0}^{R-1} h_r \left[ \sum_{m=r}^{L_w+r-1} w_{m-r} u(n-m) \right]^p.
  \label{WHmodel}
\end{equation}
After some manipulation, this relation can be put in the equivalent Volterra model form
\begin{equation*}
  y(n) = \sum_{p=1}^P \sum_{m_1=0}^{M-1} \dots \sum_{m_p=0}^{M-1} k^{(p)}(m_1,\ldots,m_p) \prod_{q=1}^p u(n-m_q),
\end{equation*}
whose symmetric discrete-time Volterra kernels are (uniquely) given by \cite{Kibangou2006}
\begin{equation}
  k^{(p)}(m_1,\ldots,m_p) = g_p \sum_{l=l_0}^L h_l \prod_{q=1}^p w_{m_q - l}, 
  \label{Vkernel}
\end{equation}
with $M=L_w+R-1$, $l_0 = \max\{0,m_1-L_w+1,\ldots,m_p-L_w+1\}$ and $L = \min\{R-1,m_1,\ldots,m_p\}$.

\subsection{CPD-based WH model identification}
\label{sec:wh-id}

We now describe the WH model identification approach proposed in \cite{Favier2009}, which involves computing the CPD of a symmetric
high-order Volterra kernel. We start by noting that, being a function of multiple discrete indices, any $p$th-order symmetric Volterra
kernel $k^{(p)}$ of memory $M$ can be uniquely identified with a $p$th-order symmetric tensor $\tenseur{X} \in \mathbb{R}^{M \times
\dots \times M}$ defined by $x_{m_1,\ldots,m_p} = k^{(p)}(m_1-1,\ldots,m_p-1)$. Owing to its convolutive form involving separable terms, the
kernel {\color{red}in} \eqref{Vkernel} can be identified with the tensor
\begin{equation}
 \tenseur{X} = g_p \sum_{r=1}^R h_{r-1} \mathbf{c}_r^{\out p} = g_p \sum_{r=1}^R h_{r-1} \left( \mathbf{S}_r \mathbf{w} \right)^{\out p},
 \label{X-CPD}
\end{equation}
where $\mathbf{S}_r \triangleq [\mathbf{e}_r \ \ \ldots \ \ \mathbf{e}_{L_w+r-1}]$, with $\mathbf{e}_m$ denoting the $m$th canonical 
basis vector of $\mathbb{R}^M$, and $\mathbf{w} = [w_0 \ \ \dots \ \ w_{L_w-1}]^T$. Expression \eqref{X-CPD} is a symmetric CPD that 
can also be written as
{\color{red}
\begin{equation}
 \tenseur{X} = \tenseur{I} \times_1 \mathbf{C} \times_2 \dots \times_{p-1} \mathbf{C} \times_p \left[g_p \mathbf{C} \diag(\mathbf{h})  
             \right],
 \label{X-matrix-CPD}
\end{equation}
where $\mathbf{h} = [h_0 \ \ \dots \ \ h_{R-1}]^T$ and $\mathbf{C} = [\mathbf{c}_1 \ \ \dots \ \ \mathbf{c}_R] \in \mathbb{R}^{M 
\times R}$}. Note that the choice of which factor is postmultiplied by $\diag(\mathbf{h})$ is irrelevant, due to the scaling 
indeterminacy. We thus conclude that the WH model \eqref{WHmodel} has equivalent symmetric Volterra kernels whose CPD are constituted 
by circulant factors $\mathbf{C}$ and a factor of the form $g_p \mathbf{C} \diag(\mathbf{h})$, which absorbs the scaling coefficients 
$g_p$ and  $h_r$.

{\color{red}
As the factors in \eqref{X-matrix-CPD} contain the parameters of the linear blocks of the WH model \eqref{WHmodel},
the above observations suggest the following three-step procedure for its identification: 
(i) estimate $k^{(p)}$ from an available set of input/output samples, using some Volterra kernel identification method (as, 
\textit{e.g.}, \cite{Kibangou2010}), (ii) compute the structured CPD from the associated symmetric tensor $\tenseur{X}$ and (iii) 
estimate the coefficients $g_q$, $q \neq p$, in the least-squares sense as explained in \cite{Favier2009}. Note that 
this requires choosing some $p \ge 3$, since otherwise the model is not identifiable: for $p=1$, it is a vector containing sums of 
products of coefficients $g_p$, $h_r$ and $w_l$; for $p=2$, we have a bilinear decomposition, which is only unique under restrictive 
assumptions (such as orthogonality). Henceforth, we assume that (i) has been accomplished and focus on 
step (ii).
}

\section{Analytical CRB for CPD-based WH estimation algorithms}
\label{sec:CRB}

\subsection{Formulation of estimation problem}
\label{sec:problem}

Let us consider that a $p$th-order tensor has been constructed from {\color{red}a non-null} estimated kernel $k^{(p)}$, as described 
in the previous section. In practice, it is evident that such a tensor satisfies
$\tenseur{Y} = \tenseur{X} + \tenseur{N}$,
where $\tenseur{N}$ is an error tensor accounting for the inevitable uncertainties which arise in the data-driven kernel estimation
procedure. Furthermore, since $k^{(p)}(m_1,\ldots,m_p)$ is symmetric in $m_1,\ldots,m_p$, in practice one estimates only the elements whose
indices pertain to a suitable non-redundant domain such as $D = \{(m_1,\ldots,m_p) : m_1 \le \dots \le m_p \}$, determining the others by
symmetry. Hence, $\tenseur{Y}$ and $\tenseur{N}$ are also $p$th-order symmetric tensors, containing redundant elements. Introducing the
selection matrix $\mathbf{\Psi} \in \mathbb{R}^{I \times M^p}$, where $I = |D| = \binom{M+p-1}{p}$, which contains as rows\footnote{The
ordering of the rows of $\mathbf{\Psi}$ is of no consequence for our purposes.} every product of the form $\mathbf{e}^T_{m_p} \kron 
\dots \kron \mathbf{e}^T_{m_1}$ for $(m_1,\ldots,m_p) \in D$, we can write the
(non-redundant) vectorized model
\begin{equation*}
 \mathbf{y} \triangleq \mathbf{\Psi} \vect(\tenseur{Y}) = \mathbf{x} + \mathbf{n} \in \mathbb{R}^I,
\end{equation*}
where $\mathbf{x} = \mathbf{\Psi} \vect(\tenseur{X})$ and $\mathbf{n} = \mathbf{\Psi} \vect(\tenseur{N})$ is a random
vector. Now, from \eqref{X-CPD}, we can deduce
\begin{align}
 \nonumber
 \vect(\tenseur{X}) 
                  = & \ g_p \sum_{r=1}^R h_{r-1} \left( \mathbf{S}_r \mathbf{w} \right)^{\kron p}
                  = \left[ g_p \sum_{r=1}^R h_{r-1} \mathbf{S}_r^{\kron p} \right] \mathbf{w}^{\kron p}  \\
                  = & \left[ g_p \sum_{r=1}^R h_{r-1} \mathbf{\Phi}_r \right]  \mathbf{w}^{\kron p}
                  = \mathbf{\Phi}(\mathbf{h}) \mathbf{f}(\mathbf{w}),
 \label{vec-X}
\end{align}
where $\mathbf{\Phi}(\mathbf{h})$ is given by the term between brackets, in which $\mathbf{\Phi}_r = \mathbf{S}_r^{\kron p}$, and
$\mathbf{f}(\mathbf{w}) = \mathbf{w}^{\kron p}$.

Our problem can therefore be expressed as that of estimating the parameters $g_p$, $\mathbf{w}$ and $\mathbf{h}$ of the WH model from
observations which satisfy $\mathbf{y} = \mathbf{\Psi} \mathbf{\Phi}(\mathbf{h}) \mathbf{f}(\mathbf{w})  + \mathbf{n}$. We assume that the
random vector $\mathbf{n}$ has zero-mean i.i.d.~components drawn from the Gaussian distribution with variance $\sigma^2$.

\subsection{Identifiability}
\label{sec:identifiability}

Due to the inherent scaling indeterminacy of our model, its local identifiability is only guaranteed with further assumptions. To
eliminate this indeterminacy, {\color{red}we assume $w_0=g_p=1$, which is sufficient due to the model structure}. Note that this 
entails no loss of generality, as {\color{red}$\mathbf{h}$ and the other coefficients $w_l$ can be rescaled accordingly.} 
Defining now $\tilde{\mathbf{w}}$ such that $\mathbf{\mathbf{w}} = [1 \ \ \tilde{\mathbf{\mathbf{w}}}^T]^T$, we can write the parameter 
vector of the WH model as $\boldsymbol \eta = [\tilde{\mathbf{w}}^T \ \ \mathbf{h}^T]^T \in \mathbb{R}^{M}$.
Global identifiability, on the other hand, is related to the uniqueness of the structured CPD. As the $k$-rank \cite{Kolda2009, 
Lim2014} of $\mathbf{C}$ equals $R$, uniqueness follows from Kruskal's condition \cite[Sec.~3.2]{Kolda2009} if $\|\mathbf{h}\|_0 = R$ 
(which implies that the $k$-rank of $\mathbf{C}\diag(\mathbf{h})$ is $R$) and $R\ge 2$. If $\|\mathbf{h}\|_0 < R$, then the $k$-rank 
of $\mathbf{C}\diag(\mathbf{h})$ equals zero; in this case, Kruskal's condition is only met for $P=4$ if $R\ge3$ and for $P\ge5$ if 
$R\ge2$.

\subsection{Parameter estimation algorithms}
\label{sec:algorithms}

In this section, we briefly review two methods that can be used to estimate the parameters $\boldsymbol \eta$ of a model of the form
\eqref{X-CPD}. \vspace{-1.5ex}

\subsubsection{Circulant-constrained ALS algorithm}
\label{sec:CALS}

The first method consists of a specialization of the well-known alternating least squares (ALS) algorithm in which the factor matrices of
the CPD are constrained as in {\color{red}\eqref{X-matrix-CPD}}. In the case of a CPD involving only circulant factors, such strategy 
has already been followed in \cite{Goulart2014}, leading to the CALS algorithm. Here, we adapt that algorithm for our purposes.

Initially, we define $\mathbf{E}_l \triangleq [\mathbf{e}_l \ \ \dots \ \ \mathbf{e}_{R+l-1}] \in \mathbb{R}^{M \times R}$, for $l \in
\{1,\ldots,L_w\}$, and $\mathbf{E} \triangleq \left[ \text{vec}\left( \mathbf{E}_1 \right) \ \ \dots \ \ \text{vec}\left( \mathbf{E}_{L_w}
\right) \right]$. With these definitions, we have 
$\vect(\mathbf{C}) = \mathbf{E} \mathbf{w}$. Next, we note that any flat matrix unfolding of $\tenseur{Y}$ can then be written as
$$~\hspace{-1em}
\mathbf{Y} \!\approx\!   \ \mathbf{C} \diag(\mathbf{h}) \left( \mathbf{C}^{\khatri p-1} \right)^T \!=\!  \ \unvect_R  \left( \mathbf{E} \mathbf{w} \right) \diag(\mathbf{h}) \left( \mathbf{C}^{\khatri p-1} \right)^T
$$
where the above approximation  is due to the presence of noise and the operator $\unvect_R$ is defined such that, 
{\color{red}$\forall$} $\mathbf{a} = [\mathbf{a}_1^T \ \ \dots \ \ \mathbf{a}_R^T]^T$ with $\mathbf{a}_r \in \mathbb{R}^{N}$, 
$\unvect_R(\mathbf{a}) =
[\mathbf{a}_1 \ \ \dots \ \ \mathbf{a}_R]$.
{\color{red}Using}
the property $\vect(\mathbf{A}\diag(\mathbf{b})\mathbf{D}) = (\mathbf{D}^T \odot \mathbf{A}) \mathbf{b}$, we have also
$\vect(\mathbf{Y}) \approx  \left( \mathbf{C}^{\khatri p} \right) \mathbf{h}$.
Hence, given current estimates $\hat{\mathbf{w}}^{k}$ and $\hat{\mathbf{h}}^{k}$, we can update them with the scheme
\begin{align*}
   \text{(i)} \ & \hat{\mathbf{v}}^{k+1} =  \frac{1}{R} \mathbf{E}^T \vect
     \left\{\mathbf{Y} \left( \hat{\mathbf{W}^{k}}^T \right)^\dagger
           \left[ \diag\left(\hat{\mathbf{h}}^{k}  \right) \right]^{-1} \right\}, \\
   \text{(ii)} \ & \hat{\mathbf{w}}^{k+1} =  \frac{1}{\left[\hat{\mathbf{v}}^{k+1}\right]_1} \, \hat{\mathbf{v}}^{k+1}, \\
   \text{(iii)} \ & \hat{\mathbf{h}}^{k+1} =
   \left( {\hat{\mathbf{W}}^{k+1}} \odot \hat{\mathbf{C}}^{k+1} \right)^\dagger  \vect(\mathbf{Y}),
\end{align*}
where  $\hat{\mathbf{C}}^{k} = [\mathbf{S}_1 \hat{\mathbf{w}}^{k} \ \ \dots  \ \ \mathbf{S}_R \hat{\mathbf{w}}^{k}]$ and
$\hat{\mathbf{W}}^k = (\hat{\mathbf{C}}^{k})^{\khatri p-1}$. Note that, to derive {\color{red} (i), we have used 
$\mathbf{E}^\dagger = (1/R) \, \mathbf{E}^T$.}

As stopping criteria, one can check {\color{red}whether} the relative difference between two consecutive values of the 
reconstruction error
{\color{red}$
 J^{k}_{\tenseur{Y}} = \left\|\tenseur{Y} - \tenseur{I} \times_1 \hat{\mathbf{C}}^{k} \times_2 \dots
              \times_p \hat{\mathbf{C}}^{k} \diag\left(\hat{\mathbf{h}}^{k}\right) \right\|_F^2
$}
falls below some fixed threshold $\epsilon_{\tenseur{Y}} > 0$ or a maximum number of iterations $K_{\text{max}}$ is attained.
\vspace{-1ex}

\subsubsection{CPTOEP algorithm}
\label{sec:CPTOEP}

Since the objective is multimodal, the main goal is to find a good approximation of the solution by using a  
\textit{low-complexity} algorithm. In \cite{Sorensen2013}, non-iterative procedures have been proposed, which are able to compute the 
\textit{exact} CPD when matrix factors are banded or structured. Consider a matrix unfolding of $\tenseur{Y}$ under the form:
$\tilde{\mathbf{Y}}\approx (\mathbf{C}^{(1)}\khatri\mathbf{C}^{(2)})\mathbf{A}^{T}$, where the structure of 
{\color{red} $\mathbf{A} = \mathbf{C}^{(3)} \khatri \dots \khatri \mathbf{C}^{(p)}$ }
is ignored, and where $\mathbf{C}^{(n)}$ are assumed Toeplitz circulant of same size  $M\times R$, that is, they can each be expressed 
in the orthonormal basis  $\{\mathbf{E}_{\ell}, 1\le \ell\le L_w\}$ defined in Section \ref{sec:CALS}:  
\vspace{-1.5ex plus 0.5ex}
{\color{red}
 \begin{equation*}
   \mathbf{C}^{(n)} = \sum_{\ell=1}^{L_w} c^{(n)}_{\ell} \mathbf{E}_{\ell}, \qquad n \in \{1,\ldots,p\}.
 \end{equation*}}
Let $\tilde{\mathbf{Y}}=\mathbf{U}\mathbf{\Sigma}\mathbf{V}^T$ denote the SVD of $\tilde{\mathbf{Y}}$. Then there exists a matrix 
$\mathbf{N}$ such that $\mathbf{U}\mathbf{N}=(\mathbf{C}^{(1)}\khatri\mathbf{C}^{(2)})$ and 
$\mathbf{N}^{-1}\mathbf{\Sigma}\mathbf{V}^T=\mathbf{A}^T$. Following the lines of \cite{Sorensen2013}, one can find matrix 
$\mathbf{N}$ and coefficients $Z_{ij}=c^{(1)}_ic^{(2)}_j$ by solving a linear system of  $M^2R$  equations in  $L_w^2+R^2-1$  unknowns. 
If there are more equations than unknowns and if the system has full rank $R$, the solution $(\mathbf{N},\mathbf{Z})$ is unique. 
First, coefficients $c^{(1)}_i$ and $c^{(2)}_j$ are obtained from the best rank-1 approximation of matrix $\mathbf{Z}$, which 
eventually yields estimates $\hat{\mathbf{C}}^{(1)}$ and $\hat{\mathbf{C}}^{(2)}$. Next, we calculate 
$\hat{\mathbf{C}}=(\hat{\mathbf{C}}^{(1)}+\hat{\mathbf{C}}^{(2)})/2$, and the estimate of $\mathbf{h}$ is obtained as in stage (iii) of 
the CALS algorithm.

The algorithm described above is suboptimal for several reasons: (a)~the model is noisy, (b)~the {\color{red}$p$} factor matrices are 
assumed to be independent, whereas they are not, and (c)~the structure of {\color{red}$\mathbf{A}$} is ignored. Hence the 
solution obtained will be inaccurate, but can be easily refined by a quasi-Newton algorithm, as will be subsequently shown.

\subsection{Closed-form expression for the CRB}
\label{sec:crb-deriv}

If we assume that $\boldsymbol\eta$ contains deterministic parameters associated with a system of interest, we have that the 
(vectorized)
measured kernel satisfies
$
 \mathbf{y} \sim \mathcal{N} (\mathbf{x}, \sigma^2 \mathbf{I}_I),
$
where $\sigma^2$ denotes the variance of the elements of $\mathbf{n}$. Hence, the mean-square error (MSE) of any locally unbiased 
estimator
$\hat{\boldsymbol \eta}(\mathbf{y})$ satisfies
\begin{equation*}
 \label{CRB}
        E\left\{ \left\|\boldsymbol\eta - \hat{\boldsymbol \eta}(\mathbf{y}) \right\|^2\right\} \ge  \underbrace{\sum_{k=1}^{L_w-1} 
\text{CRB} \left( \tilde{w}_k \right) +\sum_{r=1}^R \text{CRB} \left( h_r \right)}_{\text{trace}(\mathbf{B}(\boldsymbol \eta))},
\end{equation*}
where the CRB matrix $\mathbf{B}(\boldsymbol \eta)$ can be computed by applying the Slepian-Bangs formula, which yields 
\cite{Boizard2015}
\begin{equation*}
 \label{Slepian}
 \mathbf{B}(\boldsymbol \eta) = \sigma^2 \left( \mathbf{J}(\boldsymbol \eta)^T \mathbf{J}(\boldsymbol \eta) \right)^{-1},
\end{equation*}
where $\mathbf{J}(\boldsymbol \eta) \in \mathbb{R}^{I \times M}$ is the Jacobian matrix given by
\begin{equation*}
 \label{jacob}
 \mathbf{J}(\boldsymbol\eta) = [\mathbf{J}(\mathbf{\tilde{\mathbf{w}}}) \ \ \mathbf{J}(\mathbf{h})]
 = \left[ \frac{\partial \mathbf{x}}{\partial \tilde{\mathbf{w}}} \ \ \frac{\partial \mathbf{x}}{\partial
             \mathbf{h}}\right].
\end{equation*}

From \eqref{vec-X} and the definition of $\mathbf{f}$, we have
\begin{align*}
 \frac{\partial \mathbf{x}}{\partial \tilde{\mathbf{w}}}
       = \mathbf{\Psi} \mathbf{\Phi}(\mathbf{h}) \frac{\partial \mathbf{f}}{\partial \tilde{\mathbf{w}}}
       = \mathbf{\Psi} \mathbf{\Phi}(\mathbf{h}) \left[ \mathbf{z}_1(\tilde{\mathbf{w}}) \ \ \dots \ \
                    \mathbf{z}_{L_w-1}(\tilde{\mathbf{w}})\right],
\end{align*}
in which $\mathbf{z}_l(\tilde{\mathbf{w}}) =  \sum_{q=1}^p \mathbf{w}^{\kron q-1} \kron \mathbf{e}_{l+1} \kron \mathbf{w}^{\kron p-q}$
(with the convention $\mathbf{w}^{\kron 0} = 1$). To derive $\mathbf{J}(\mathbf{h})$, we first apply the property
$\vect(\mathbf{A}\mathbf{B}\mathbf{D}) = (\mathbf{D}^T \kron \mathbf{A}) \vect(\mathbf{B})$ to write
\begin{equation*}
   \mathbf{x}
         =  \vect(\mathbf{\Psi}\mathbf{\Phi}(\mathbf{h})\mathbf{f}(\mathbf{w}))
         =  \left( \mathbf{f}^T(\mathbf{w}) \kron \mathbf{\Psi} \right) \vect(\mathbf{\Phi}(\mathbf{h})),
\end{equation*}
leading thus to
\begin{equation*}
 \label{Jacob-lambda}
 \frac{\partial \mathbf{x}}{\partial \mathbf{h}} = \left( \mathbf{f}^T(\mathbf{w}) \kron \mathbf{\Psi} \right)
                     \left[ \vect(\mathbf{\Phi}_1) \ \ \dots \ \ \vect(\mathbf{\Phi}_R) \right].
\end{equation*}

In order to identify the contribution of $\mathbf{w}$ and $\mathbf{h}$ in ${\rm CRB}(\tilde{w}_k)$ and ${\rm CRB}({h}_r)$, we propose 
to extend the results presented in~\cite{Boizard2015} by using oblique projection. This is the purpose of the following proposition.
{\color{red}We denote by} $\mathbf{E}_{\mathbf{A} \mathbf{B}}$ the oblique projection whose range is $\langle \mathbf{A} \rangle$ and 
whose null space contains $\langle \mathbf{B} \rangle$ (see~\cite{scharf94a} for details).

\begin{proposition}
The closed-form expression for the CRB of $\tilde{{w}}_k$ is given by: \vspace{-1ex}
$$
{\rm CRB}(\tilde{w}_k) = \frac{\sigma^2}{\| \mathbf{g}_{k} \|^2 -   \| \mathbf{E}_{\mathbf{G}_{k} \mathbf{J}(\mathbf{h})} 
\mathbf{g}_{k} \|^2 -   \| \mathbf{E}_{\mathbf{J}(\mathbf{h}) \mathbf{G}_{k} } \mathbf{g}_{k}\|^2},
$$
where $\mathbf{g}_k$ is the $k$th column of $\mathbf{J}(\tilde{\mathbf{w}})$ and $\mathbf{G}_k$ is the submatrix of 
$\mathbf{J}(\tilde{\mathbf{w}})$ obtained by removing its $k$th column.
Similarly, the closed-form expression for the CRB of $h_r$ is:
$$
{\rm CRB}({h}_r) = \frac{\sigma^2}{\| \mathbf{d}_{r} \|^2 -   \| \mathbf{E}_{\mathbf{D}_{r} \mathbf{J}(\tilde{\mathbf{w}})} 
\mathbf{d}_{r} \|^2 -   \| \mathbf{E}_{\mathbf{J}(\tilde{\mathbf{w}}) \mathbf{D}_{r} } \mathbf{d}_{r}\|^2},
$$
where $\mathbf{d}_r$ is the $r$th column of $\mathbf{J}({\mathbf{h}})$ and $\mathbf{D}_r$ is the submatrix of 
$\mathbf{J}({\mathbf{h}})$ obtained by removing its $r$th column.
\end{proposition}
The proof is omitted due to the lack of space.

\section{Simulation results}
\label{sec:simulation}

To illustrate the utility of the derived CRB, we now present some Monte Carlo simulation results. Specifically, we evaluate
several estimators when applied to identify a WH model with parameters $\mathbf{w}^T = [$1 0.538 1.834 -2.259 0.862$]^T$, $\mathbf{h}
= [$1.594 -6.538 -2.168$]^T$ from estimates of the equivalent symmetric third-order kernel $\tenseur{X}$, proceeding as follows. For
each realization of the (symmetric) noise tensor $\tenseur{N}$, we vary $\sigma^2$ and then construct a data tensor
$\tenseur{Y}=\tenseur{X}+\tenseur{N}$  for each chosen level of $\sigma^2$. Next, we compute estimates {\color{red}$\hat{\boldsymbol
\eta}(\mathbf{y}) \in \mathbb{R}^7$} given by:
 (i) the family of estimators $N$-CALS, 
 which consist in applying $N$ times the algorithm of Section \ref{sec:CALS} with random initializations and keeping the best solution in
terms of reconstruction error (w.r.t.~$\tenseur{Y}$);
 (ii) the estimator CPTOEP, described in Section \ref{sec:CPTOEP};
 (iii) the estimator CPTOEP-CALS, which corresponds to refining the CPTOEP estimate by applying the CALS algorithm;
 (iv) the estimator CPTOEP-BFGS, in which a similar refinement is obtained by minimizing a least-squares criterion (w.r.t.~$\tenseur{Y}$)
with the Broyden--Fletcher--Goldfarb--Shanno (BFGS) algorithm\footnote{Specifically, we used the Fortran implementation whose Matlab
interface is available at \tt{http://github.com/pcarbo/lbfgsb-matlab}.} \cite{Byrd1995}.
The maximum number of iterations established for CALS and BFGS is $K_{\text{max}} = 2000$. We choose $\epsilon_{\tenseur{Y}} = 10^{-10}$
and set the tolerance of BFGS also {\color{red}as} $10^{-10}$.
For each estimate $\hat{\boldsymbol\eta}(\mathbf{y})$, we compute $\varepsilon_{\boldsymbol \eta} = \|\boldsymbol \eta - \hat{\boldsymbol
\eta}(\mathbf{y})\|^2$. This procedure is repeated for 100 realizations of $\mathcal{N}$ and then $\varepsilon_{\boldsymbol\eta}$ is
averaged for each level of $\sigma^2$, yielding a mean-square error estimate denoted by MSE$_{\boldsymbol\eta}$.

\begin{table}[t]
\caption{Simulation results: estimated MSE$_{\boldsymbol\eta}$ values (in dB).}
\label{tab:scen-1}
\centering
\vskip2mm
\footnotesize
\renewcommand{\arraystretch}{1}
\tabcolsep=0.15cm
 \begin{tabular}{c|c|c|c|c|c|c}
 \cline{2-7}
  & \multicolumn{6}{c}{$1/\sigma^2$ (dB)} \\
  \hline
Estimator &      10 &     20 &     30 &     40 &     50 &     60 \\
\hline
 1-CALS &  19.22 &  17.14 &  18.37 &  17.68 &  18.53 &  17.86 \\
 5-CALS & -15.04 & -25.05 &   4.04 &   4.05 & -55.07 &   4.06 \\
10-CALS & -15.04 & -25.05 & -35.04 & -45.07 & -55.02 & -65.07 \\
CPTOEP & -13.96 & -23.94 & -33.94 & -43.94 & -53.94 & -63.94 \\
CPTOEP-CALS & -15.04 & -25.04 & -35.04 & -45.05 & -55.02 & -65.13 \\
CPTOEP-BFGS & -20.04 & -30.03 & -40.03 & -50.02 & -60.01 & -69.62 \\
\hline
CRB & -20.18 & -30.18 & -40.18 & -50.18 & -60.18 & -70.18 \\

\hline
 \end{tabular}
\end{table}

The results are shown in Table \ref{tab:scen-1}, as well as the computed values of the CRB. One can see that 1-CALS has a very poor
performance, due to its frequent premature termination or inability to converge. Although 5-CALS performs better, its results are
degraded for the same reasons. CPTOEP, in its turn, performs slightly worse than 10-CALS, but attains a similar level when refined by
CALS. Yet, there remains a gap between their MSE curves and that of the CRB. Indeed, only {\color{red}CPTOEP-BGFS} attains an MSE
close to the CRB. Note that a similar gap has been reported by \cite{Liu2001} for the ALS algorithm. Along the lines of their 
discussion, we believe that, in the case of CALS, this gap is due to the convergence problems which are always observed in practice, 
at least for a few runs. As for CPTOEP, this seems to happen because the adapted procedure yields suboptimal estimates.

Finally, we note that the above comparison is justified since, under the assumption of Gaussian additive noise, the least-squares criterion
leads to the maximum likelihood (ML) estimator. In signal-in-noise problems, the ML estimator is often approximately unbiased even for a
small sample size, provided that the SNR is sufficiently high \cite{Kay1993}.

\section{Conclusion}
\label{sec:conclusion}

A closed-form expression of the CRB has been derived for the parameter
estimates of a CPD having {\color{red} identical banded circulant factors, one of which is post-multiplied by a diagonal 
scaling matrix}. Then, two specialized algorithms have been proposed to compute a CPD {\color{red}with that structure}. The 
first{\color{red}, named CALS,} is an adaptation of the ALS method taking the structural constraints into account, whereas the second 
is composed of two steps: (i) compute an approximate solution thanks to a non iterative algorithm (CPTOEP), and (ii) refine the 
solution {\color{red}via CALS or via} a quasi-Newton descent (BFGS). The latter (CPTOEP-BFGS) reached the Cram\'er-Rao bound over a 
wide range of SNR values. The proposed algorithms have been applied to identify {\color{red}a WH model}, and their statistical 
performance has been evaluated using the derived CRB.




\end{document}